# Multi-scale filler structure in simplified industrial nanocomposite silica/SBR systems studied by SAXS and TEM

Guilhem P. Baeza[1,2,3], Anne-Caroline Genix[1,2], Christophe Degrandcourt[3], Laurent Petitjean[3], Jérémie Gummel[4], Marc Couty[3], Julian Oberdisse[1,2]

[1] *Université Montpellier 2, Laboratoire Charles Coulomb UMR 5221, F-34 095, Montpellier, France*

[2] *CNRS, Laboratoire Charles Coulomb UMR 5221, F-34095, Montpellier, France*

[3] *Manufacture Française des Pneumatiques MICHELIN, Site de Ladoux, 23 place des Carmes Déchaux, F-63 040 Clermont-Ferrand, Cedex 9, France*

[4] *European Synchrotron Radiation Facility, ESRF, 6 rue Jules Horowitz, BP 220, F-38 043, Grenoble, Cedex 9, France*

10/12/12

**Abstract**

Simplified silica (Zeosil 1165 MP) – SBR (140k carrying silanol end-groups) nanocomposites have been formulated by mixing of a reduced number of ingredients with respect to industrial applications. The thermo-mechanical history of the samples during the mixing process was monitored and adjusted to identical final temperatures. The filler structure on large scales up to microns was studied by transmission electron microscopy (TEM) and very small angle X-ray scattering (SAXS).

A complete quantitative model extending from the primary silica nanoparticle (of radius $\approx 10$ nm), to nanoparticle aggregates, up to micron-sized branches with typical lateral dimension of 150 nm is proposed. Image analysis of the TEM-pictures yields the fraction of zones of pure polymer, which extend between the branches of a large-scale filler network. This network is compatible with a fractal of average dimension 2.4 as measured by scattering. On smaller length scales, inside the branches, small silica aggregates are present. Their average radius has been deduced from a Kratky analysis, and it ranges between 35 and 40 nm for all silica fractions investigated here ($\Phi_{si} = 8 - 21\% v$).

A central piece of our analysis is the description of the inter-aggregate interaction by a simulated structure factor for polydisperse spheres representing aggregates. A polydispersity of 30% in aggregate size is assumed, and interactions between these aggregates are described with a hard core repulsive potential. The same distribution in size is used to evaluate the polydisperse form factor. Comparison with the experimental intensity leads to the determination of the average aggregate compacity (assumed identical for all aggregates in the distribution, between 31% and 38% depending on $\Phi_{si}$), and thus aggregation number (ca. 45, with a large spread). Due to the effect of aggregate compacity and of pure polymer zones, the volume fraction of aggregates is higher in the branches than $\Phi_{si}$. The repulsion between aggregates has a strong effect on the apparent isothermal compressibility: it leads to a characteristic low-q depression, which cannot be interpreted as aggregate mass decrease in our data. In addition, the reinforcement effect of these silica structures in the SBR-matrix is characterized with oscillatory shear. Finally, our results show that it is possible to analyze the complex structure of interacting aggregates in nanocomposites of industrial origin in a self-consistent and quantitative manner.

**Figures: 9**

**Tables: 1**

# 1. Introduction

The mechanical reinforcement of polymer matrices by nanoparticles is a fundamental problem with far reaching applications, e.g., for car tires [1, 2]. From a conceptual point of view, it is generally recognized that the filler structure has a strong impact on the mechanical properties [3-5], accompanied by the effect of chain structure evolving in the hard filler environment [6-9], and the filler-chain interactions [10-15]. All these contributions are related to the filler structure, and it is thus important to be able to characterize it in detail. Unfortunately, two typical situations are usually encountered: either the system is a model system of individually dispersed nanoparticles [16], which is easier to understand but is further away from applications, or the system is made by mixing of powders of aggregated nanoparticles, together with many additives, and analysis becomes difficult. In the literature, the list of typical ingredients of industrial systems comprises the filler and the polymer matrix (often styrene-butadiene rubber (SBR), or polybutadiene), silane coupling agents like TESPT, known also as Si69, or its successor Si363, coating or compatibilizing agents like octyl-triethoxysilane (octeo) with catalyzers (diphenyl guanidine (DPG)), cross-linking agents (sulfur), cure activators like ZnO nanoparticles, stearic acids, accelerator providing fast cure rate like sulfenamides (TBBS, CBS), antioxidants like various substituted paraphenylene diamines (PPD) and phenol-based antioxidants (AO2246).[17-21] Note that ZnO nanoparticles, e.g., may contribute to the scattering signature even at low concentrations due to their high electron density [22], unless their contribution is suppressed using sophisticated anomalous scattering techniques [23]. In contrast, apart from the antioxidants added after the polymer synthesis, we address here the issue of a simplified industrial system containing only the filler and its compatibilizer octeo with DPG. We have compiled typical industrial formulations in the appendix highlighting the reduction of parameters in our simplified system.

Structural studies of the dispersion of precipitated silica filler of the type used here have been undertaken by several groups. Ramier et al have studied the silica structure in SBR by transmission electron microscopy (TEM) and small-angle x-ray scattering (SAXS), without further analysis of the SAXS-data as they focused on the rheology.[20] Conzatti et al have investigated the morphology of the same silica in SBR by TEM with automated image analysis, and dynamic mechanical analysis (DMA), varying the surface modification.[18] A similar approach was presented by Stöckelhuber et al on the flocculation of precipitated silica as a function of coupling.[21] The reinforcement by fractal aggregates with again the same silica in SBR was addressed by Mélé et al by SAXS and AFM.[19] An in-depth analysis of the SAXS-data was outside the scope of this article. Other groups have focused on the fractal dimensions extracted from the power-law decay of the scattered intensity.[24, 25] Schneider et al have presented scattering data on precipitated silica in poly(dimethylsiloxane) and SBR, with a two-level description based on the Teixeira [26] or Beaucage [27] equations for fractals made of beads.[28, 29] A qualitative analysis of SAXS-curves has been proposed by Shinohara et al.[30] In the present paper, a quantitative view in the same spirit will be presented. Several theoretical studies on scattering in complex systems have been published. For example, Schweizer et al on interacting fractals propose apparent structure factors as a function of filler volume fraction.[31]

Analysis of structural data is usually considerably less difficult and ambiguous in model systems. In such systems, the filler particles are available as individually dispersed beads in a solvent, and particular care is taken by the experimentalists in order to maintain or control colloidal stability throughout the nanocomposite formulation process, which is often solvent casting. Meth et al have studied silica nanoparticles in poly(methyl methacrylate) and polystyrene (PS) by SAXS.[32] Some aggregation is often present and visible at small angles, but due to the high monodispersity bead-bead interaction peaks are found. Janes et al have investigated the influence of annealing history on the structure as seen by scattering.[33] After strong annealing, no low-angle indication for aggregation is found and the curves strongly resemble perfect dispersions of spheres. In articles by Chevigny et al [15, 34] and Jouault et al [35, 36], the structure of silica nanoparticles in PS is analyzed, by TEM and scattering. There, a focus is made on the relationship between filler structure and mechanical reinforcement in systems with well-defined dispersion of small aggregates in the matrix. Reverse Monte Carlo (RMC) modeling has been applied to interacting aggregates measured in a silica–latex model system.[37] Recently, we have also contributed with a structural model used to follow the film formation of silica-latex films.[8]

The effect of fractal aggregates on the rheology has been investigated theoretically in a seminal paper by Witten et al [38]. Several empirical equations usually based on extensions of the original Einstein hydrodynamic reinforcement [39, 40] exist and have been summarized in the literature [41]. The standard analysis is commonly based on three methods: either DMA, where the sample is subject to oscillatory torsion at fixed amplitude and frequency, as a function of temperature, e.g. [42, 43]. From such studies, carried further with NMR measurements, a strong interest in the so-called glassy layer of polymer on filler surface has arisen.[10, 44, 45] The second standard method is oscillatory linear rheology in shear, which however is limited to low moduli [35, 46]. The latter one is often applied to characterize the non linear Payne-effect at small strains, which is responsible for the decrease of the storage modulus with shear amplitude [20, 21, 47-49]. The third method is uniaxial strain, which is strongly non linear and leads to high deformations, up to rupture [50, 51]. The latter method is often used to characterize the Mullins effect [52]; it can also be combined with scattering [53], or with NMR as done in an outstanding paper by Klüppel et al in SBR nanocomposites [54]. In many of the rheological studies of nanocomposite systems, the effect of e.g., silane coupling or compatibilization is studied and tentatively correlated with filler structure, if available.[20, 47] For the sake of completeness, other techniques allowing a characterization of the dynamics of nanocomposite systems are dielectric spectroscopy [55, 56] and quasi-elastic neutron scattering [57-60].

In this article, we investigate a 'simplified industrial system', i.e., ingredients have been limited to the strict minimum. For the structural analysis of nanocomposites of increasing filler fraction, we have chosen to combine direct imaging methods like TEM, which has the advantage of intuitive analysis but the drawback of limited representativity of local details, with a reciprocal space method, SAXS, which is highly representative but is difficult to interpret. The structure of the silica within the nanocomposites will be modelled in a step-by-step, multi-scale manner, starting with the primary silica beads as basic building units (10 – 20 nm range). These beads are found to be aggregated in small clusters, the typical radius of which (40 nm range) will be determined by Kratky plots. These aggregates are themselves concentrated in large-scale fractal branches (thickness ca. 150 nm, extending over microns). Inside these branches, the small aggregates repel each other. Within our model, this is described with a hard-sphere excluded volume interaction potential, which induces a characteristic depression of the scattered intensity at intermediate angles. This depression is directly related to the local concentration of aggregates, which is higher than the nominal silica volume fraction due to the confinement in the fractal branches, and the presence of polymer inside the aggregates. Therefore, a quantitative TEM analysis was used to estimate the volume fraction of fractal branches, $\Phi_{fract}$. Secondly, we have set up an independent Monte Carlo (MC) simulation in order to calculate the low-q limit of the polydisperse inter-aggregate structure factor, which quantifies the depression. Using a polydisperse aggregate form factor obeying the same polydispersity, the mass of the small aggregates (or, equivalently, their internal silica volume fraction, here called compacity) and their concentration inside the fractal branches can be extracted from the scattered intensity. In parallel, mechanical measurements allowed us to extract an average aggregate compacity in good agreement with the former analysis.

The outline of this article is as follows. After the materials and methods section, all results are discussed in section 3. The thermomechanical history of the mixing process characterized by the observed torque and temperature during mixing is discussed in section 3.1. It is followed by an analysis of the large-scale structure of the nanocomposites using TEM and the low-angle scattering in section 3.2. The next section is devoted to an in-depth analysis of the complete scattering curve, which takes aggregation and interaction between aggregates into account. Finally, the rheological and mechanical properties are studied and discussed in section 3.4.

## 2. Materials and methods

**Nanocomposite formulation:** Silica-SBR nanocomposites are formulated by stepwise introduction and mixing of SBR chains with silica pellets in an internal mixer (Haake). Note that particular care was taken to avoid any trace of carbon black, catalysing nanoparticles (ZnO), crosslinking or coupling agents, which may impede interpretation of, e.g., scattering experiments. Compared to the complex samples usually studied in the literature, our system is thus designed to be a simplified industrial nanocomposite, conserving namely aggregated multi-scale silica as filler particles, SBR-chains, and a mixing process, all related to tire applications.

The mixing chamber is preheated as a function of nanocomposite composition, in order to obtain the same final mixing temperature of 160±5°C, and thus a comparable thermo-mechanical history. For the same reason, the rotor speed is adjusted during the process to between 95 and 105 rpm. The polymer is introduced first, in the form of centimetric lamellae. After about one minute, the mixture of silica pellets, DPG (Vulcacit from Bayer, 1%w with respect to polymer), and the liquid coating agent (octeo from Dynasylan, 8%w with respect to silica) is incorporated via the same piston. The process is finished after typically five minutes. The hot sample is then rapidly cooled and homogenized by lamination 10 times in the 1 mm gap between rotating cylinders (two roll mill). The silica volume fractions in the nanocomposites reported here have been measured by thermogravimetric analysis (Mettler Toledo) using a first ramp at 30 K/min from 25°C to 550°C under nitrogen, followed by a second ramp at 20 K/min from 550°C to 750°C under air. They are found to be systematically by 15% lower than the weighted quantities, presumably due to losses in the mixer. Only the silica volume fractions above 8%v have been considered here. For lower silica contents, inhomogeneous composites were obtained due to a less effective mixing process.

**System characterization:** The silica pellets (Zeosil 1165 MP from Rhodia) have been dispersed by sonification in water under basic conditions, and have been studied by SAXS and small-angle neutron scattering (SANS). The resulting scattering pattern is rather unstructured, indicating high polydispersity. A characteristic size corresponding to a radius of 10 nm is found. A complete analysis reveals a lognormal size distribution ($R_0 = 8.55$ nm, $\sigma = 27\%$, leading to $<R_{si}> = 8.9$ nm and an average bead volume of $V_{si} = 3.6 \ 10^3$ nm$^3$, the corresponding specific surface is 160 m$^2$/g), in agreement with TEM studies. $V_{si}$ will be used to estimate aggregation numbers of silica nanoparticles (or beads) in aggregates.

The polymer (with antioxidants 6PPD and AO2246) has been purpose-synthesized by Michelin, and the chain mass characterized by size exclusion chromatography. The polymer matrix is made of two types of chains of molecular mass 140 kg.mol$^{-1}$ (PI = 1.07). Each chain is a statistical copolymer with styrene (26%w) and butadiene (74%w) units (41% of which are 1-2 and 59% of 1-4). The glass-transition temperature as measured by differential scanning calorimetry (DSC, 200F3 Maia from Netzsch) with a heating rate of 20 K/min is -31°C. This is in agreement with Fox's law predicting -38°C for a mixture of 1,2- and 1,4-butadiene and styrene ($T_g$(1,4-butadiene 59%) = -65°C [61]; $T_g$(PS) = 100°C) and suggests that polymerization is thus indeed statistical. For the loaded samples, $T_g$ shows no significant variation as compared to the pure SBR matrix: $T_g = -32 \pm 0.5$°C for all the silica contents investigated here. 50% of the chains are linear unmodified SBR-chains, whereas the other 50% bear a single graftable silanol end-function. This functional group may graft the chain on the silica by silanol condensation with the surface silanol.

**Structural analysis:** The silica microstructure in the nanocomposites has been studied by transmission electron microscopy and SAXS. TEM pictures were obtained with samples prepared by ultracryomicrotomy at -80°C on a LEICA FC-7 (Diatome ultra 35°, desired thickness set to 50 to 70 nm). Electron microscopic observations in transmission were achieved with a Philips CM200 LaB6 (200 kV, bright field mode). A grey-scale analysis of the pictures using ImageJ was performed to determine the pure polymer fraction. The average and the standard deviation of this quantity were obtained via a statistical analysis over several pictures (e.g., 12 for the sample with 8.4%v of silica). SAXS experiments (beamline ID2, ESRF, Grenoble) were performed at a wavelength of 1.1 Å (12.46 keV), using two sample-to-detector distances (1 m and 10 m), yielding a total q-range from 0.001 to 0.5 Å$^{-1}$. Even lower-q data was measured on the Bonse-Hart set-up on ID2 ($q_{min} = 10^{-4}$ Å$^{-1}$). Samples were cut into pieces of approximate thickness 0.8 mm. The scattering cross section per unit sample volume $d\Sigma/d\Omega$ (in cm$^{-1}$) – which we term scattered intensity I(q) – was obtained by using standard procedures including background subtraction and calibration given by ESRF. The contrast of silica in polymer in SAXS experiments was calculated from the scattering length densities ($\rho_{SBR} = 8.85 \ 10^{10}$ cm$^{-2}$, $\rho_{si} = 1.97 \ 10^{11}$ cm$^{-2}$, $\Delta\rho = 1.09 \ 10^{11}$ cm$^{-2}$), which were themselves known from the chemical composition.

**Data analysis of small-angle scattering:** The scattering patterns of industrial nanocomposites usually show a complex multi-scale behaviour. Starting at high q (q > 1/$R_{si}$), the signature of the primary particles can be found, and in particular their specific surface, associated with a particular scattering power law. When going towards intermediate q, a break in slope (or peak) may be observed. Its position, $q_{si}$, is related to the typical interparticle distance, and for crowded nanoparticles in contact, it is located close to $\pi/R_{si}$. If a superstructure of nanoparticles exists, then another break in slope at low q-values may be found, located at the inverse of the typical size of such structures. For aggregates in contact, e.g., the position is $q_{agg} = \pi/R_{agg}$. To summarize this overview, different scaling regimes may be observed, the transition between them identifying characteristic sizes. It is possible to visualize these breaks in slope by counterbalancing the power-law decrease. In Kratky-plots, for instance, $q^2I(q)$ is plotted versus q. The breaks in slope then appear as easily recognizable maxima.

The relationship between the cross-overs of the scaling laws, and the typical sizes characterizing the microstructure, can be clarified based on the general scattering law for spherically symmetric, monodisperse particles and aggregates, which is presented for didactical reasons first. A generalization to include polydispersity will be developed afterwards, cf. the Monte Carlo simulations below. Besides the difference in contrast $\Delta\rho$, the scattered intensity I(q) of monodisperse and spherical silica beads in the polymer can be written in an identical manner for SAXS and SANS:

$$I(q) = \Phi_{si} \, \Delta\rho^2 \, V_{si} \, S(q) \, P(q) \tag{1}$$

where $\Phi_{si}$ is the volume fraction of silica, $\Delta\rho$ the contrast between silica and the SBR matrix, $S(q)$ the total bead-bead structure factor, and $P(q)$ the normalized form factor of the beads, i.e. $P(q{\rightarrow}0) = 1$.

If the silica particles are organized in aggregates of approximately the same size, the total structure factor $S(q)$ may be factorized in two terms [62]: the inter-aggregate structure factor $S_{inter}(q)$, which is the Fourier transform of the pair-correlation function of the center-of-mass of (average) aggregates. The second is the intra-aggregate structure factor, $S_{intra}(q)$, which is the Fourier transform of the pair correlation of beads inside the aggregate. If a higher order organization of such aggregates exists, a third structure factor $S_{fract}(q)$ describing this structural level may be introduced to describe the complete q-range:

$$S(q) = S_{fract}(q) \, S_{inter}(q) \, S_{intra}(q) \tag{2}$$

Even in absence of any detailed structural model, the fractal structure factor of non interacting fractals has the following properties: At low q, it decreases from the total mass of the fractal with a characteristic power law, $\sim 1/q^d$, where d is the fractal dimension of the network. At higher q, which is where $S_{inter}(q)$ begins to describe the intermediate scale structure, $S_{fract}$ equals one. The location of cross-over from one regime to the other can be estimated, e.g., based on a fractal made of monodisperse spheres of radius a. The cross-over is then located at $q_{branch} = \sqrt{3}/(e \, a)$, where e is the Euler constant, i.e., the lateral branch dimension is $2a \approx 1.3/q_{branch}$. Another property of this structure factor is that it can be approximated by the sum of 1 (at intermediate and high q) and a low-q power law. This is the reason why low-q power laws can in general be *subtracted* from the total scattered intensity in spite of the *product* in eq.(2). The product of $S_{intra}(q)$ and the form factor of the particles can then be merged into a single expression, form factor of the average aggregate:

$$P_{agg}(q) = S_{intra}(q) \, P(q) \tag{3}$$

The limiting value of $P_{agg}$ at low q is given by the aggregation number, $N_{agg}$. At high q, $S_{intra}$ tends to one, and $P_{agg}$ thus reproduces the local structure of the beads making up the aggregates. The transition of $S_{intra}(q)$ from $N_{agg}$ (and some power law) at low q to one at high q is achieved typically at $q_{si}$, where one may also find a structure factor peak in case of close contact. The internal structure described by $S_{intra}(q)$ is thus the origin of the first break in slope discussed above. The multiplication by the other factors in eq. (1) may slightly shift this feature. The same argument may then be repeated on a bigger scale with $S_{inter}(q)$, which is responsible of (at least) one break in slope at lower angles. Finally, note that the low-q limit of the structure factor tends towards the (relative) isothermal compressibility. In systems with repulsive (e.g., hard core) interactions, this compressibility may be well below one, and therefore decrease the low-q scattering. This decrease is a concentration effect which cannot be interpreted as a reduction of the aggregate mass.

**Monte-Carlo simulation of the structure factor of polydisperse systems:** We have discussed structure factors in monodisperse systems in the preceding section. In this article, a fully polydisperse description will be used. In this case, the relevant structure factor, $S_{inter}$, has to be replaced by an apparent one, $S_{inter}^{app}$, the calculation of which is outlined here. We have used a simulation box containing between 20 000 and 50 000 beads depending on the volume fraction in the range from 5%v to 30%v in order to have a roughly constant box size, $L_{box} = 2.5$ μm. The minimum accessible q value is obtained from $55/L_{box} \approx 2.8 \ 10^{-3}$ Å$^{-1}$ (the prefactor of 55 has been determined by comparison with the known monodisperse Percus-Yevick structure factor). Here we are interested in the effect of polydispersity in aggregate size on the low-q limit, which is why the exact radius of the bead representing the aggregate is not of importance. In our simulations, the bead radius is described by a lognormal size distribution for the polydispersity with $R_0 = 20$ nm and $\sigma = 0$, 15%, and 30%. This size distribution has been converted in a 15-population histogram. Standard Monte-Carlo steps verifying the excluded volume conditions have been performed. After equilibration, the partial structure factors $S_{ij}(q)$ between populations i and j have been calculated using the Debye formula [63]. They were used to determine the total scattered intensity which is given as the sum of the products of $S_{ij}(q)$ weighted by the appropriate form factors $P_i(q)$ and $P_j(q)$. The apparent structure factor is then obtained by dividing by the average form factor:

$$S_{inter}^{app}(q) = \frac{I(q)}{I_0 P(q)} = \frac{\sum_{i,j} \sqrt{N_i N_j} \ V_i V_j \ \sqrt{P_i(q)P_j(q)} \ S_{ij}(q)}{\sum_i N_i \ V_i^2 \ P_i(q)} \qquad (4)$$

**Rheology:** The rheological response in the linear regime of the nanocomposites was obtained with a stress-controlled rheometer AR 2000, used in the strain-controlled mode in plate-plate geometry (20 mm diameter). Isothermal frequency sweeps at fixed low deformation level ($\gamma =$ 0.1%) were performed in the temperature range from 10°C to 80°C. Using the principle of time-temperature superposition, master curves of the storage modulus, G'(ω), and the loss modulus, G"(ω), corresponding to measurements at 50°C were established between ω = 2πf = 2π 10⁻³ and 2π 10³ rad/s.

## 3. Results and discussion

### 3.1 Thermo-mechanical characterization of the mixing process

During the mixing process in the internal mixer, the torque as a measurement of the evolving material viscoelasticity, and the temperature are monitored as a function of time. These quantities are plotted in Figure 1a and 1b, respectively, for various silica volume fractions from 8.4%v to 21.1%v. The incorporation of silica and octeo leads to an important increase in torque after about 1min30. During the nanocomposite mixing, the silica pellets are crushed for several minutes. This leads to the desired temperature increase to about 160°C (Figure 1b), which is essential for the silanol end-function (50% of reactive chains) and octeo grafting chemistry. Towards the end of the mixing, the high temperature induces a decrease in nanocomposite viscosity, as is visible in Figure 1a. As expected, a higher silica volume fraction leads to a higher maximum torque.

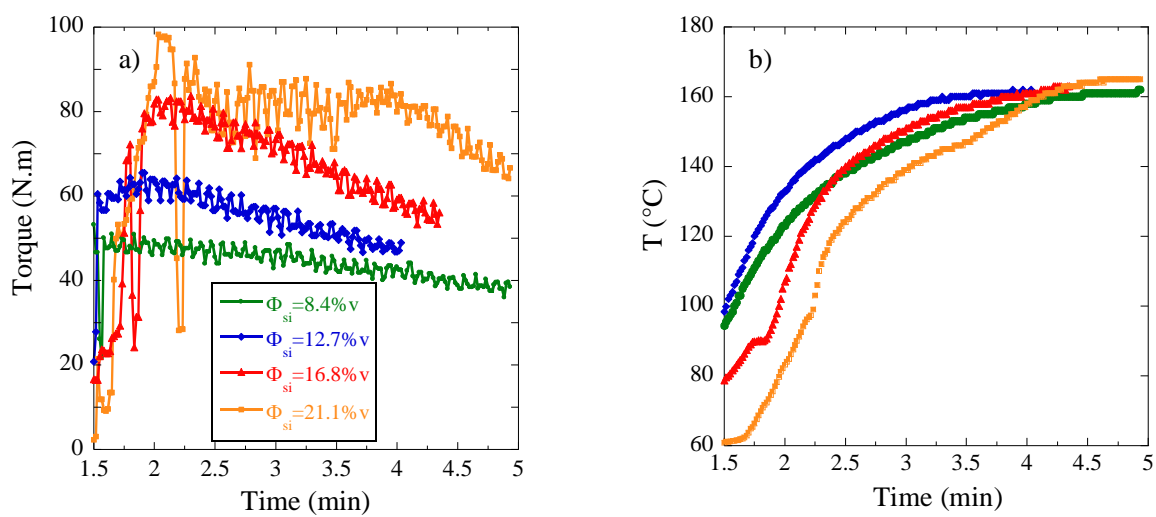

**Figure 1**: **(a)** Torque observed during mixing of SBR nanocomposites for a series in silica volume fractions (8.4%v – 21.1%v). **(b)** Temperature in the mixing chamber of the same samples during the process.

### 3.2 Large-scale structural characterization by SAXS and TEM

The microstructure of the silica in nanocomposites has been studied by x-ray scattering and transmission electron microscopy. The scattering data are shown in Figure 2a for the series in silica volume fractions in the SBR matrix. If one wishes to compare samples of different silica contents, it is obvious from eq.(1), that the reduced representation $I(q)/\Phi_{si}$ gives direct access to the variations in the structure factor S(q), P(q) being fixed.

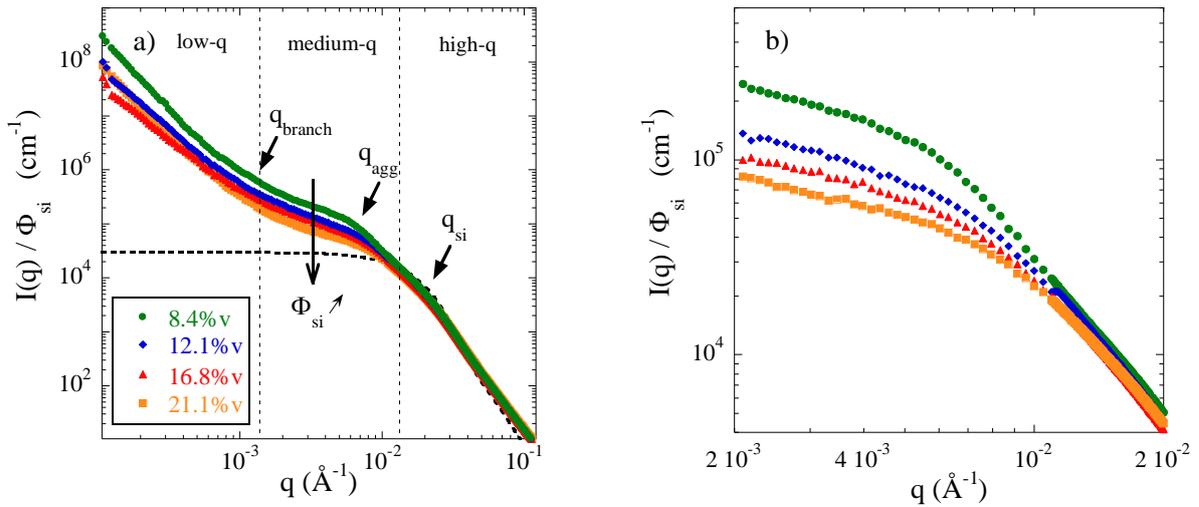

**Figure 2:** Silica structure of nanocomposites studied by SAXS. **(a)** Reduced scattered intensity $I(q)/\Phi_{si}$ for a series of silica volume fractions in matrix (8.4% v - 21.1% v). Dotted line: form factor of the silica beads. Arrows indicate the breaks in slope discussed in the text. **(b)** Intermediate-q structures highlighted after subtraction of the low-q power law.

In Figure 2a the complete scattering curves are shown. There is a strong low-q upturn at q-values down to $10^{-4}$ Å$^{-1}$. It cannot be described by any simple function, but can roughly be represented by a power law $Aq^{-d}$, with fractal dimension $d = 2.4 \pm 0.3$. It can also be noted that in the reduced representation $I(q)/\Phi_{si}$, the value of the prefactor A decreases with increasing $\Phi_{si}$. We will see below that this is related to the decrease in isothermal compressibility at intermediate q-values.

The standard model of fractal structures made of blob-structures relates the radius of the fractal, $R_{fract}$, to the number of spherical subunits, $N_b$, and their radius, a:

$$R_{fract} = a \, N_b^{1/d} \qquad (5)$$

In Figure 2a, the transition from the large-scale network to internal branch structure, i.e., the breakdown of fractality, can be located at the intersection of the power-laws describing the low-q and the intermediate-q scattering, around $q_{branch} = 1.10^{-3} - 2.10^{-3}$ Å$^{-1}$. Using the fractal model of agglomerated spheres for the large-scale fractal structure factor, $S_{fract}(q)$, as outlined in section 2, the lateral branch dimension can be estimated to $2a = 1.3/q_{branch} \approx 100$ nm, with large error bars due to the limited precision on the crossover and the rudimentary model. We will see below that electron microscopy gives 150 nm. On the other extreme of the geometry of the fractals, in the q-range under study, there is no appreciable cut-off of the power-law at low q. This means that their upper size $R_{fract}$ extends up to the micron range. The mass-fractal model (eq.(5)) can be used to estimate the pure polymer fraction between branches, (1- $\Phi_{fract}$), where $\Phi_{fract}$ denotes the volume fraction of fractal branches. For micron-size fractals, a rough estimate of the pure polymer fraction of $\approx 84\%$ is found, which is certainly an overestimation due to the unrealistic space-filling properties of spheres as compared to branches. To summarize this analysis, the large-scale structure of the nanocomposite as seen by SAXS up to dimensions of microns can be interpreted as a network of branch size around 100 nm, and significant amounts of empty space between them.

Figure 2b focuses on the intermediate and high-q features, after subtraction of the low-q upturn. A slowly varying scattering curve is found for all silica volume fractions in the intermediate q-range. A model for the structures observed in this range will be proposed in the next section.

The large-scale structure of nanocomposites has been studied by TEM. In Figure 3, representative pictures for two samples ($\Phi_{si} = 8.4$ and $21.1\%$v) are shown.

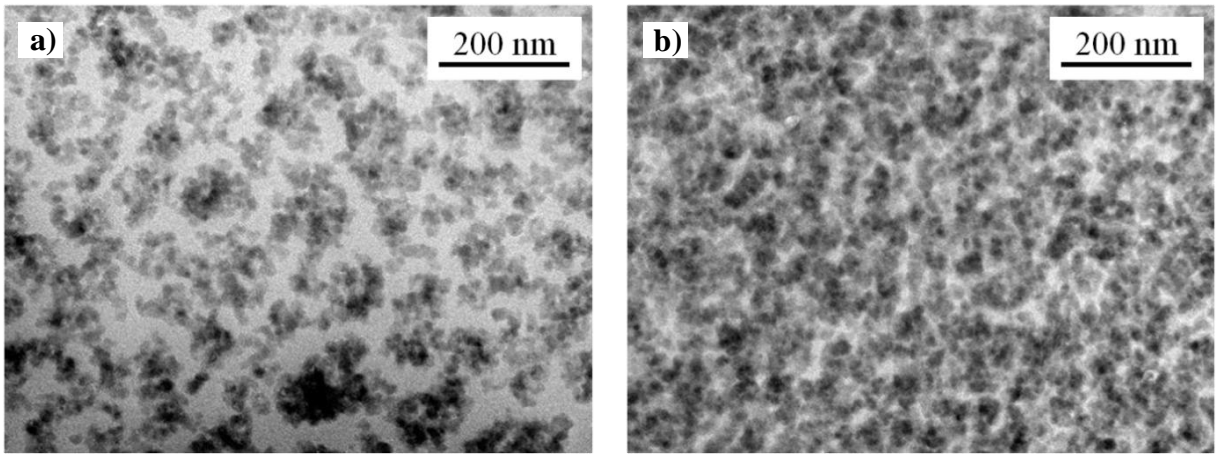

**Figure 3:** TEM-pictures of nanocomposite samples: **(a)** $\Phi_{si} = 8.4\%$v **(b)** $\Phi_{si} = 21.1\%$v.

The 8.4%v-sample shows nice structural features, which can be summarized as follows: (a) A dense branched structure of lateral dimension of around $2a \approx 150$ nm is seen, made of small silica beads aggregated together. (b) A grey-scale analysis of the pictures reveals that the pure polymer fraction is about $41 \pm 4\%$ in surface. Note that in the slice, most of the silica beads are visible individually, leading to a first level of grey, whereas a small number overlap giving a darker grey. From simple geometric considerations, it appears that in thin enough slices, of thickness smaller than the structural length under study in the sample ($\approx 150$ nm), the surface and volume fractions of matter (branches) coincide. It is thus concluded that approximately 41% of the sample is not occupied by branches. Similarly, the higher volume fraction sample shown in Figure 3b is much denser, with a pure polymer fraction of about $20 \pm 4\%$ in surface (and thus also in volume).

### 3.3 Modelling of the SAXS-data on intermediate and small scales

In this section, the average aggregate size based on a Kratky analysis, the inter-aggregate structure factor, and the aggregate form factor will be discussed. Putting these separated descriptions together will allow us to extract the average aggregate compacity, and thus aggregation number. Note that our model is based on a full description of polydispersity: aggregates monodisperse in size would lead to unphysical characteristics (namely compacity), and would contradict the TEM pictures.

**Average aggregate size.** We start with the discussion of the highest curve ($\Phi_{si} = 8.4\%$v) in Figure 2. The two high-q arrows indicate the cross-overs between three power laws, at $q_{si} = 0.022$ Å$^{-1}$ and $q_{agg} = 0.0065$ Å$^{-1}$. The ratio $q_{si}/q_{agg}$ suggests that this first superstructure has a typical linear dimension of only some four bead-sizes, which is why it is identified with small aggregates. Following our interpretation outlined in section 2, $\pi/q_{si}$ gives a typical bead radius. 14.1 nm is found, bigger than but of the same magnitude as the silica beads ($<R_{si}> = 8.9$ nm). From the second break in slope, an aggregate radius which we associate with the average $<R_{agg}> = \pi/q_{agg} = 48.4$ nm is deduced. Introducing the compacity $\kappa$, or internal aggregate volume fraction, the aggregation number $N_{agg}$ can be related to $<R_{agg}>$. Allowing for a generalization to polydispersity, the definitions for an aggregate of radius $R_{agg}$ read:

$$\kappa = \frac{V_{si\,in\,agg}}{V_{agg}} \qquad (6)$$

$$N_{agg} = \frac{4}{3}\pi R_{agg}^{3} \frac{\kappa}{V_{si}} \qquad (7)$$

$V_{agg} = 4\pi/3\,R_{agg}^{3}$ is the total volume of such an aggregate, and $V_{si\,in\,agg}$ the volume effectively occupied by silica in this aggregate, i.e., $N_{agg}V_{si}$, with $V_{si}$ the average bead volume. The aggregates (with $<R_{agg}> = 48.4$ nm) are rather small, as even if one assumes random close packing ($\kappa = 0.64$)[64], only $N_{agg} \approx 83$ beads would be part of one aggregate. The use of more realistic values for $\kappa$ (i.e., below 64%) would give lower aggregation numbers. A model for the determination of $\kappa$ will be developed below including a polydisperse description of both the aggregate form factor and the inter-aggregate structure factor.

In Figure 2b, the low-q upturn discussed in section 3.2 has been subtracted. The scattering curves at different $\Phi_{si}$ in the reduced representation $I(q)/\Phi_{si}$ differ at low q, and are seen to overlap perfectly above a critical wave vector $\approx 0.01$ Å$^{-1}$, corresponding to primary silica nanoparticles in close contact. Due to the high-q overlap, the break in slope associated with the nanoparticle size is seen to stay constant: indeed, its value is 13.7 nm for the higher silica concentrations (12.7, 16.8, and 21.1%v), and we will use an average value of 13.85 nm in the Kratky analysis below. On the other hand, the break in slope associated with aggregate size moves to higher q-values with increasing $\Phi_{si}$. The associated aggregate radius $<R_{agg}>$ decreases to 39.2 nm (resp. 36.1 and 34.4 nm) for 12.7%v (resp. 16.8%v and 21.1%v).

In order to determine the position of $q_{agg}(\Phi_{si})$ more precisely, a Kratky presentation of the data has been chosen. In Figure 4a, the breaks in slope are seen to be transformed in well-identified maxima. A multi-parameter fit of the two overlapping maxima has been achieved using the following sum of two functions $G_1$ and $G_2$:[65]

$$G_i(q) = \frac{A_i}{\sqrt{2\pi}\sigma_i q}\exp\left(-\frac{\ln^2\left(\frac{q}{q_i}\right)}{2\sigma_i^2}\right) \qquad (8a)$$

$$G(q) = G_1(q) + G_2(q) \qquad (8b)$$

Each of these functions describes a lognormal function of amplitude $A_i$, position $q_i$, and width $\sigma_i$ (i = si, agg). Note that the parameter $q_i$ is slightly shifted to higher values as compared to the peak position, but is preferred due to its vicinity with the corresponding break in slope (see in Figure 4b). Again, we associate $q_{agg}$ with $<R_{agg}>$. The width and position of the high-q lognormal describing the silica bead $q_{si} = (\pi/13.85 \text{ nm})$ was kept fixed, thereby reducing the number of free parameters. An example of the underlying lognormals is shown in the inset of Figure 4a for the 8.4%v-sample.

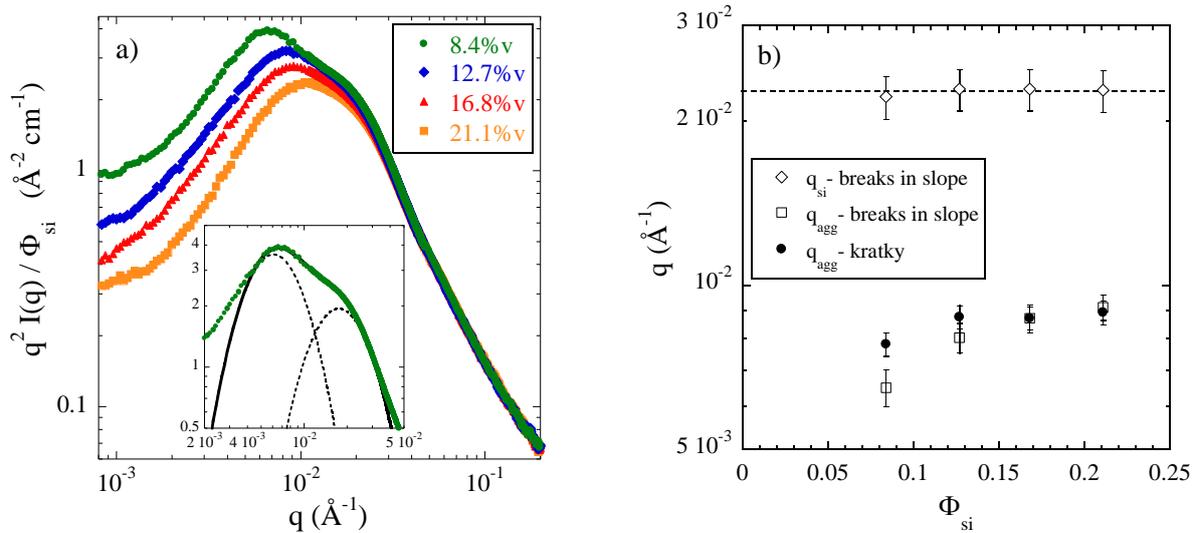

**Figure 4**: **(a)** Same data as Fig. 2 in Kratky representation of the reduced scattered intensity $q^2I(q)/\Phi_{si}$ for $\Phi_{si} = 8.4 - 21.1\%$v. Inset: Zoom on 8.4%v-data with fit by sum of two lognormal functions. **(b)** $\Phi_{si}$-dependence of the lognormal position parameters $q_{si}$ (dotted line) and $q_{agg}$ associated with the inter-aggregate structure factor. $q_{si}$ and $q_{agg}$ values obtained from the breaks in slope are also included.

The lognormal position parameters related to the aggregates are found to evolve less than the breaks in slope discussed before. All values are plotted in Figure 4b, together with the silica bead peak position fixed in the Kratky analysis. The corresponding average aggregate radii are $<R_{agg}> = 40.2$ nm (resp. 35.9, 36.1, and 35.2 nm) for 8.4%v (resp. 12.7, 16.4, and 21.1%v). To summarize, both methods of analysis – breaks in slope and Kratky plots – give similar aggregate radii, in the range between 34 and 48 nm. The maxima in the Kratky plots are better defined, and the radii grouped together, between 35 and 40 nm, values which we will use in the following analysis as the average values of the size distributions.

**Polydisperse inter-aggregate structure factor.** As already indicated above, the aggregate compacity $\kappa$ is a key quantity, as it relates the size of the aggregates to the amount of silica they carry, i.e., it characterizes the internal aggregate structure. It has also a strong impact on the overall structure of the sample: by silica volume conservation, the higher the compacity, the less aggregates are located in a given volume to satisfy the nominal volume fraction, $\Phi_{si}$. As a consequence, the number density of aggregates depends on $\kappa$ and affects the (inter-aggregate) structure factor, $S_{inter}^{app}$. Increasing the number of mutually repelling aggregates leads to a decrease in the isothermal compressibility, a feature which is clearly visible in Figure 2b: the intermediate and low-q reduced intensity decreases with increasing $\Phi_{si}$. In this picture, the Kratky-peak is due to the excluded volume interactions between aggregates, and thus located close to $q_{agg} = \pi/<R_{agg}>$. The observation of this peak together with the low-q decrease suggests two points. First, it is not possible to conclude on aggregate mass and size from a pure form factor analysis of the intensity decrease in such interacting systems. This decrease is caused by the structure factor dependence on the filler concentration. Secondly, one may quantitatively account for the decrease using a model for the structure factor of polydisperse hard spheres representing aggregates, which is what is proposed now.

In order to obtain the polydisperse structure factor, we have performed Monte Carlo simulations. In case of polydispersity in size, no general formula exists, and the thermodynamic properties of the system are not described by a single structure factor any more[66]. The partial structure factors $S_{ij}(q)$ between two size populations i and j, weighted by the form factors of these populations, have to be added up to obtain the total intensity. Often, an apparent structure factor $S_{inter}^{app}(q)$ obtained by dividing the intensity by the average form factor is used, as defined in eq.(4). It is not a thermodynamic quantity because of its dependence on the shape and contrast of the objects, but can be looked at as a useful representation of the correlations. Our approach is the following: the $S_{inter}^{app}(q)$ have been calculated by MC simulations assuming excluded volume interactions as described in section 2. They are shown in Figure 5a for different volume fractions $\Phi_{agg}$ of polydisperse spheres representing aggregates. The low-q limiting values, $S_{inter}^{app}(q \to 0)$, are needed to determine the aggregate compacity in the next section. They have been determined by extrapolation, as presented in Figure 5a. In our model, the aggregate polydispersity is the only unknown parameter. From the absence of a strong peak at close contact – only a break in slope is observed in Figure 2b – it is concluded that polydispersity of aggregates in size is at least 30%. Then the structure factor peak of, e.g., the simulation with $\Phi_{agg} = 20\%v$ is close to one, in agreement with the experimental data. In addition, taking a too low polydispersity would lead to unphysical aggregate compacities. Polydispersity of aggregates has thus been fixed to 30% in our model, i.e., of the same order as the primary bead polydispersity. Such a value also accounts for the fact that one cannot distinguish aggregates of finite size in the TEM pictures (Figure 3).

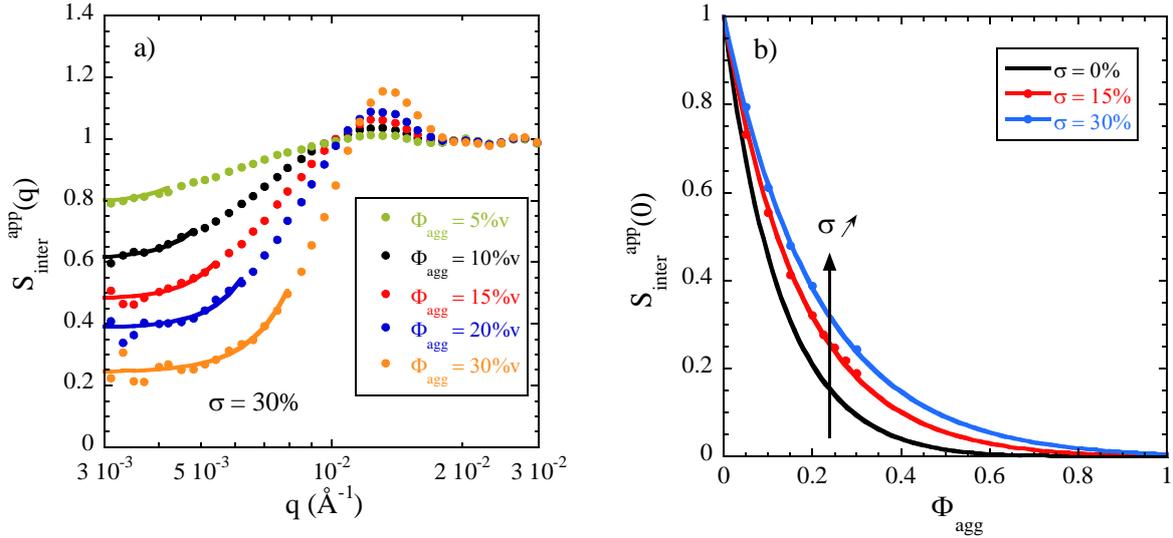

**Figure 5:** MC simulations of polydisperse systems. **(a)** Apparent structure factor obtained by the division of simulated I(q) by the average P(q) as defined in eq.(4). I(q) is calculated for polydisperse hard spheres representing the aggregates using a lognormal distribution with $R_0 = 20$ nm (arbitrarily fixed) and $\sigma = 30\%$. Lines are fit of the low-q part with an arbitrary function: $S_{inter}^{app}(q) = S_{inter}^{app}(0) + (Aq)^B$ in order to extract $S_{inter}^{app}(0)$. **(b)** Evolution of $S_{inter}^{app}(0)$ versus the aggregate volume fraction for $\sigma = 0$, 15% and 30%. Lines are fits using eq.(9).

After extrapolation, the obtained $S_{inter}^{app}(q \rightarrow 0)$ values are plotted in Figure 5b, for various polydispersities ($\sigma = 0$, 15%, 30%), as a function of $\Phi_{agg}$. The description of these values can be achieved using a Percus-Yevick (PY) structure factor [67, 68]. Its limiting value for q→0 can be easily determined from the full expression:

$$S_{PY}(q \rightarrow 0) = \frac{\left(1 - \alpha\Phi_{agg}\right)^4}{\left(1 + 2\,\alpha\Phi_{agg}\right)^2} \tag{9}$$

Here we have introduced a prefactor $\alpha$ for the volume fraction. $\alpha$ equals one for the standard monodisperse PY-formula, and will be used here as a free parameter for polydisperse spheres. Indeed, it is observed in Figure 5b that a higher polydispersity leads to higher low-q limiting values, as if the higher polydispersity had a similar effect at low-q than decreasing the volume fraction. A surprisingly good fit is found with the PY-expression given in eq.(9), with $\alpha = 0.72$ and $0.60$ for $\sigma = 15\%$ and $30\%$, respectively. This enables us to use the PY-equation with $\alpha = 0.6$, representing a typical polydispersity of $\sigma = 30\%$, for $S_{inter}^{app}(q \to 0)$ and thus the determination of the aggregate compacity $\kappa$.

To finish this discussion on the silica microstructure as probed by SAXS, $S_{inter}^{app}$ has a low-q contribution below one due to aggregate repulsion. $S_{inter}^{app}$ corresponds to the structure factor of an *infinite* homogeneous sample of aggregates at aggregate volume fraction $\Phi_{si}/(\kappa\Phi_{fract})$, whereas here aggregates are only in the branches. The point is that this apparent isothermal compressibility is lower for more concentrated samples, and by continuity this intensity depression is passed on to the structure factor describing the fractal: the complete scattering curve is thus lowered in the intermediate- and low-q range.

**Polydisperse aggregate form factor.** Our analysis is based on Figure 2b. Combining eqs.(1-3) and subtracting the low-q upturn treated in the preceding paragraph, the scattering at intermediate q reads for a polydisperse system:

$$\frac{I(q)}{\Phi_{si}} = \Delta\rho^2 V_{si} S_{inter}^{app}(q) \left\langle P_{agg}(q) \right\rangle \tag{10}$$

For $I(q \to 0)$, we focus on the region around $q^* = 0.003$ Å$^{-1}$. Such a value provides a good estimate (compared to $\pi/\langle R_{agg} \rangle$) for the determination of the low-q limit $S_{inter}^{app}(q \to 0)$, which has been calculated in the preceding section. We now focus on the average form factor of aggregates. The calculation is based on the polydispersity in radius of the aggregates. We have seen that the absence of the structure factor peak suggests a polydispersity of $\sigma = 0.3$. From the Kratky plots, the average aggregate radius was determined. For the example of $\Phi_{si} = 8.4\%v$, aggregates are chosen to be described (as in the simulation) by a lognormal distribution of radii, with parameters $R_0 = 38.4$ nm and $\sigma = 0.3$, giving the average of $\langle R_{agg} \rangle = 40.2$ nm. The conversion into aggregate mass is based on the main assumption of the polydisperse description: It is assumed that the compacity $\kappa$ is the same for all aggregates of different size. One can thus use eq. (7) to transform the size distribution in the distribution of $N_{agg}$, an example of which is shown in Figure 6, for $\kappa = 31\%$.

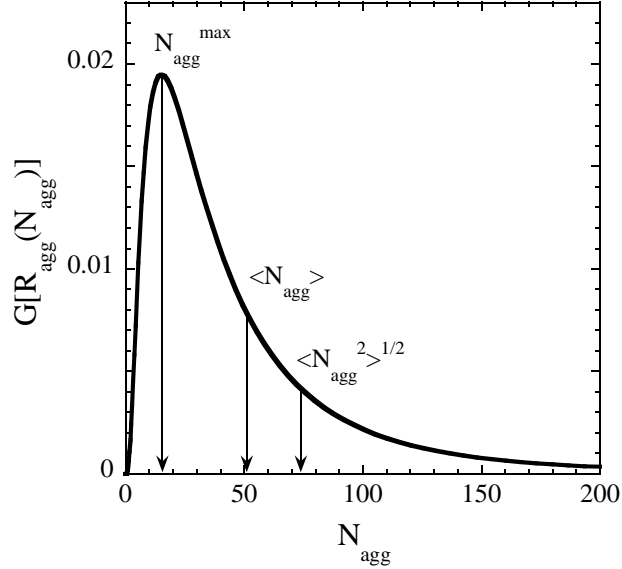

**Figure 6:** Aggregation number distribution function deduced from the lognormal distribution of radii (parameters $R_0 = 38.4$ nm, $\sigma = 0.3$) and eq.(7) supposing $\kappa = 31\%$.

Concerning the aggregate form factor, recall that in the monodisperse case, $P_{agg}(q\rightarrow0) = N_{agg}$. For polydisperse systems, $P_{agg}(q\rightarrow0) = <N_{agg}^2>/<N_{agg}>$. The moments of $N_{agg}$ are easily calculated from the distribution function (Figure 6). At non zero q (we focus on $q^* = 0.003$ Å$^{-1}$), the decrease of the form factor of the aggregates has to be included. In this limit, the polydisperse form factor becomes:

$$\left\langle P_{agg}(q*) \right\rangle = \frac{\left\langle N_{agg}^2 \right\rangle}{\left\langle N_{agg} \right\rangle} \exp\left(-\frac{q*^2 R_G^2}{5}\right) \tag{11}$$

where $N_{agg}$ depends on compacity via eq.(7) and the radius distribution function, and $R_G^2 = <R_{agg}^8>/<R_{agg}^6>$ is the correctly averaged Guinier radius[69].

**Determination of compacity.** The description of the scattered intensity (eq.(10)) includes both previously defined quantities, the structure factor $S_{inter}^{app}$ and the average form factor $<P_{agg}>$ (eq.(11)). The apparent isothermal compressibility $S_{inter}^{app}(q \to 0)$ depends on the aggregate volume fraction in the branches, which is given by

$$\Phi_{agg} = \frac{\Phi_{si}}{\Phi_{fract} \, \kappa} \qquad (12)$$

The volume fraction of fractal $\Phi_{fract}$ appears because it accounts for the concentration effect in the fractal branches, due to the existence of pure polymer zones surrounding the branches. The latter quantity has been determined by TEM in section 3.2 for the highest and lowest $\Phi_{si}$ values, and interpolated in between.

The procedure to determine the compacity $\kappa$ (assumed to be identical for all aggregates in the distribution) is thus to (a) assume an initial value for $\kappa$, (b) calculate the structure factor with eqs.(9) and (12), (c) determine the distribution of $N_{agg}$, (d) calculate $<P_{agg}>$ using eq.(11), and (e) assess the intensity level (eq.(10)). $\kappa$ is then changed until eq.(10) for $q = q^*$ is fulfilled. The values of $\kappa$ for all silica volume fractions are reported in Table 1, together with aggregate radii, average aggregation numbers, width of dispersion, and radius of an aggregate of average aggregation number.

| $\Phi_{si}$ | $<R_{agg}>$ (nm) $\pm 5\%$ | $\Phi_{fract}$ $\pm 4\%$ | $\kappa$ $\pm 2\%$ | $\Phi_{agg}$ $\pm 3\%$ | $<N_{agg}>$ $\pm 15\%$ | $\Delta N_{agg}$ | $R_{agg}^{eq}$ (nm) $\pm 5\%$ |
|---|---|---|---|---|---|---|---|
| 8.4% v | 40.2 | 0.59 | 0.31 | 0.47 | 51 | 53 | 52.3 |
| 12.7% v | 35.9 | 0.66 | 0.33 | 0.57 | 40 | 43 | 46.9 |
| 16.8% v | 36.1 | 0.73 | 0.36 | 0.64 | 44 | 47 | 47.1 |
| 21.1% v | 35.2 | 0.80 | 0.38 | 0.69 | 44 | 47 | 46.0 |

**Table 1:** Results of the analysis of SAXS data of nanocomposites containing polydisperse aggregates ($\sigma = 0.3$ in radius). Average aggregate radius $<R_{agg}>$ (from Kratky analysis), volume fraction of fractal branches $\Phi_{fract}$, compacity $\kappa$, aggregate volume fraction $\Phi_{agg}$, average aggregation number $<N_{agg}>$, standard deviation of the distribution in $N_{agg}$, and equivalent radius of an aggregate of average mass.

Following this procedure, the compacity found for, e.g., $\Phi_{si} = 8.4\%v$ is 31%. The resulting distribution function of $N_{agg}$ was already shown in Figure 6. One immediately sees in this figure that the $R_{agg}{}^3$-dependence strongly increases the polydispersity and asymmetry of the distribution, which has a pronounced tail. The resulting average of $N_{agg}$ is 51, and the standard deviation $\Delta N_{agg} = \sqrt{(<N_{agg}{}^2> - <N_{agg}>^2)} = 53$, i.e., of the same order of magnitude, which illustrates the width of the distribution. For comparison, the radius of the average aggregate ($<N_{agg}> = 51$) is also given in the table ($R_{agg}{}^{eq} = 52.3$ nm). On the other hand, most of the aggregates are considerably smaller, as the peak of the distribution is located close to $N_{agg}{}^{max} \approx 15$ (corresponding to $R_{agg} \approx 35$ nm). Again, such a wide distribution is in line with the TEM observations, where actually no aggregates are clearly identified, presumably due to the large size distribution.

As the silica concentration is increased, only minor changes are observed in Table 1: clusters conserve their average mass (in the range between $<N_{agg}> = 40$ and 51), but contract slightly, leading to an increase of their compacity from 31 to 38%. Note that such compacities are compatible with the choice of hard sphere interactions for the inter-aggregate structure factor. Concentrating aggregates in the fractal branches induces a considerable depression of the scattering (via the isothermal compressibility) at intermediate q-values, as observed in Figure 2. Again, in the light of the discussion including both $S_{inter}^{app}(q)$ and $<P_{agg}(q)>$ in eq. (10), interpreting the intensity decrease at intermediate-q as a decrease of $<P_{agg}>$ only could lead to the erroneous interpretation of decreasing aggregate mass. On the contrary, our analysis confirms that the average aggregate mass remains approximately constant in our system. Finally, one may note that the error bar on $\Phi_{fract}$ has only a minor effect ($\pm 5\%$) on $<N_{agg}>$, whereas the 5% error on $<R_{agg}>$ causes most of the uncertainty on $<N_{agg}>$ ($\pm 15\%$).

In Figure 7, the real-space and reciprocal space models in terms of the three structure factors are shown. In real space, the multi-scale structure is represented by the large scale fractal network of dimension 2.4, the branches of which are made of dense assemblies of aggregates of typical aggregate radius, $R_{agg}$, and volume fraction $\Phi_{agg}$. Finally, these aggregates are themselves made up of on average some forty-five primary particles of radius $R_{si}$, and possess a compacity $\kappa$, which is typically 35%.

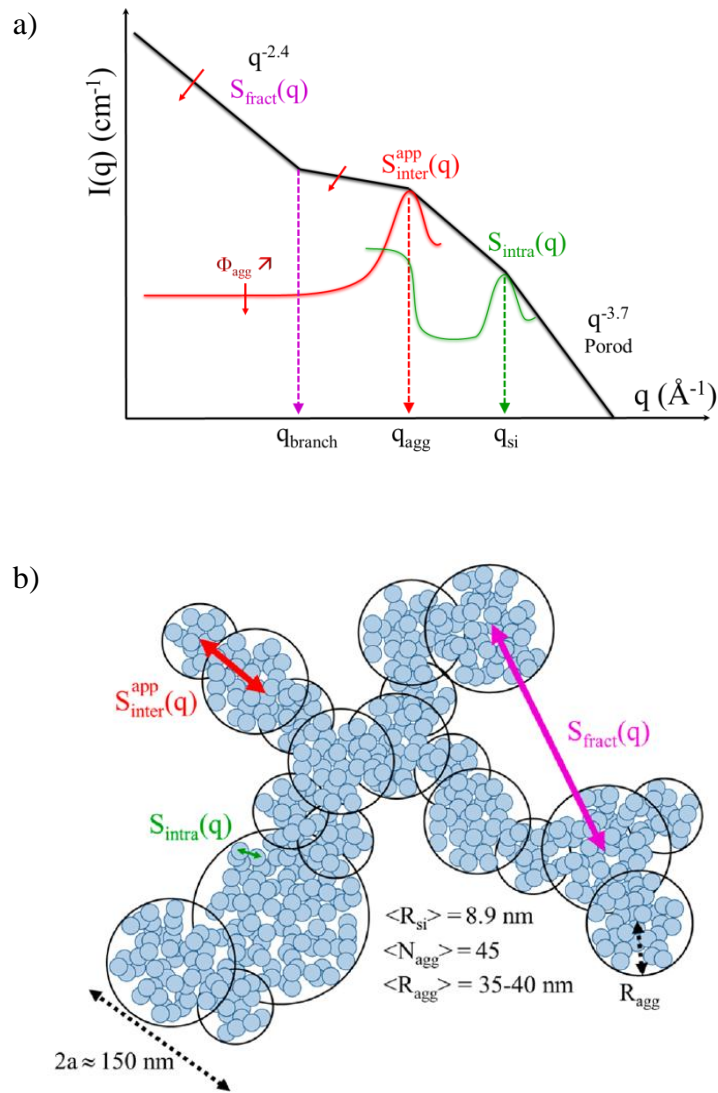

**Figure 7:** **(a)** Model decomposition of the scattered intensity into the three contributions to the reciprocal-space structure: $S_{fract}(q)$, $S_{inter}^{app}(q)$ (red line), and $S_{intra}(q)$ (green line, $S_{intra}(q \rightarrow 0) = \langle N_{agg}^2 \rangle / \langle N_{agg} \rangle$). **(b)** Real-space representation of the multi-scale structure of the silica nanocomposites associated with **(a)**.

### 3.4 Rheology and reinforcement

The motivation for the determination of the structure of simplified industrial nanocomposites relies in its link with the remarkable rheological and mechanical properties of such materials. Therefore, we focus now on the characterization of the rheology of these nanocomposites. In absence of curing agents in our simplified formulation, nanocomposites are not crosslinked. Silica-free samples are thus polymer melts, i.e. viscoelastic liquids; adding silica may change their rheological character. The series of samples of increasing silica volume fraction (0 – 21.1% v) has been studied by small amplitude (linear regime) oscillatory shear experiments. Moduli at various temperatures have been measured and used for the construction of master curves for G'($\omega$) and G''($\omega$) applying the time-temperature superposition at a reference temperature of 50°C. At low enough $\Phi_{si}$ ($\leq$ 12.7% v), the superposition of curves at different temperatures (horizontal shift factors are discussed below) yields unambiguous master curves. The resulting moduli of the matrix and the two lower silica volume fractions are plotted in Figure 8a. Note that no vertical shift factors were required to achieve superposition as occasionally necessary for composites[46, 70].

In the viscoelastic response of the matrix, the flow regime at low frequency (G''$\sim \omega^{1.2}$, close to the expected exponent of one), a characteristic time given by the maximum of G'' ($\tau = 1/\omega_{max} = 2$ s), and a high frequency modulus ($G_0 = 0.79$ MPa) can be identified. In addition, the G'' curve displays a high-frequency upturn towards the glass transition regime. With 8.4% v and 12.7% v of silica, respectively, the curves are mainly shifted to higher moduli: this increase will be analyzed in terms of the reinforcement factor. In parallel, small changes in the shape of the curve can be observed: the characteristic G'' maximum shifts to slightly lower frequencies with respect to the matrix, and G' and G'' overlap and finally do not cross any more at low $\omega$. A common criterion for liquid-like samples is that G'' is greater than G' in a given frequency range. The impact of filler is to increase the elastic component above the viscous one over the whole range, and thus "gel" the samples. In this case, there is no hint of a terminal relaxation for the polymer, but a solid-like behavior. This is probably related to the existence of a percolated network microstructure which is not able to relax completely and becomes more pronounced with increasing $\Phi_{si}$.

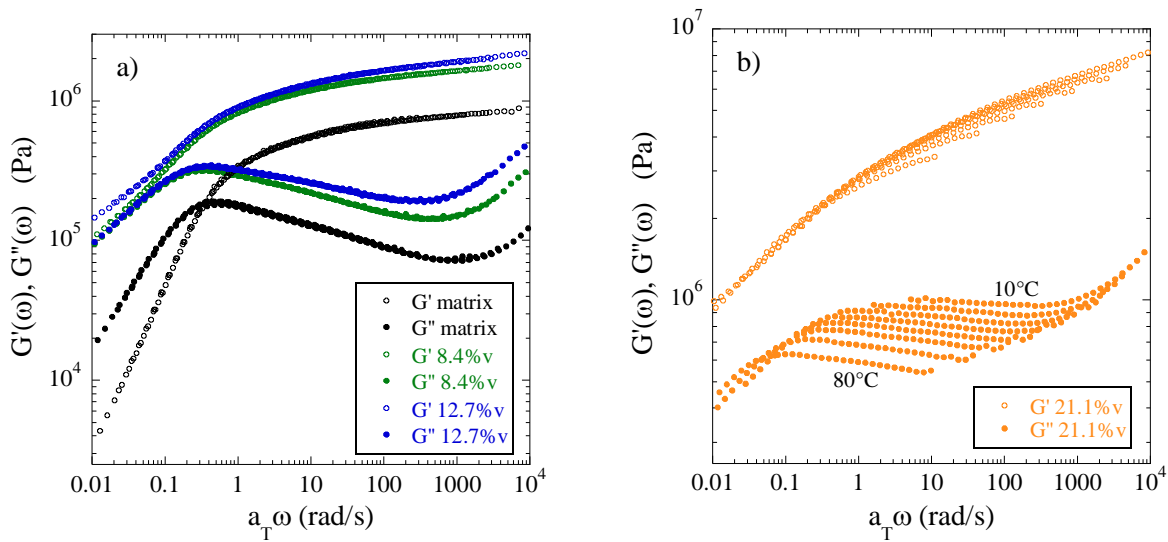

**Figure 8:** Master curves for G' and G'' (Pa) as a function of angular frequency ω (rad/s) at the reference temperature of 50°C for nanocomposites with **(a)** $\Phi_{si} = 0$, 8.4%v, 12.7%v, and **(b)** $\Phi_{si} = 21.1$%v.

As the volume fraction is increased to 16.8%v and 21.1%v, the quality of the master curves suffers. This is exemplified in Figure 8b, where in particular the G'' cannot be superimposed neatly any more by applying only horizontal shift factors. The storage moduli G' stay within an envelope, but it is unclear if these values are entirely trustworthy. The reasons for this discrepancy may lie either in slip on the plate due to the too high moduli, or in the failure of time-temperature superposition for dynamically heterogeneous samples[10]. With our data, we are unable to judge. It can be concluded that the moduli of these samples increase considerably, and that G' is always at least a factor of two higher than G''. We observe a broadening of the G'' peak in its high frequency range corresponding to a larger and more asymmetric distribution of relaxation times. It may be envisaged as a slower contribution (possible glassy layers) from the glass transition process located at higher frequency (out of our experimental window).

The horizontal shift factors, $a_T$, obtained from the master curve construction are found to change slightly from the matrix to the two lower silica contents. Their evolution with the inverse of temperature can be well described with an Arrhenius equation. The flow activation energy thus obtained is estimated to be 53, 58 and 60 kJ.mol$^{-1}$ for silica loadings of 0, 8.4 and 12.7%v, respectively. Alternatively, the classical Williams–Landel–Ferry (WLF) equation [71] could also be used leading to the system constants $C_1 = 6.9$ and $C_2 = 265$ K for the pure polymer matrix at the reference temperature of 50°C. For the nanocomposites, the values are $C_1 = 8.1$ and $C_2 = 280$ K ($C_1 = 6.1$, $C_2 = 210$ K) for 8.4%v and 12.7%v of silica, respectively. The fact that the characteristics of the time-temperature superposition ($a_T$, $E_a$ or $C_1$, $C_2$) are not significantly modified by the introduction of filler in spite of strong variation of the G' and G'' shapes was already observed in the literature for nanocomposite systems.[36, 46, 70, 72] Such results suggest that the temperature-dependent relaxation process probed here is similar in the composites and the unfilled polymer.

From the high-frequency storage modulus (estimated here at 150 Hz), the relative reinforcement of the nanocomposites $G/G_0$ with respect to the pure matrix can be calculated as a function of silica volume fraction. For the highly loaded samples, the modulus at 150 Hz is estimated from the average of the data point dispersion (see in Figure 8b). All resulting reinforcement factors are plotted in Figure 9. The reinforcement factor has the advantage of highlighting the influence of the filler, as it cancels the matrix contribution. It can also be compared directly to the Einstein equation for hydrodynamic reinforcement [39], and its application by Smallwood [40] or Mooney [73] to reinforcement of polymer matrices (see, e.g., ref [41] for different reinforcement factor descriptions). Here, a specific model based of percolation of aggregates inside the branches, which themselves extend across the whole sample, is proposed. Indeed, Figure 8a suggests a cross-over from liquid-like to solid-like behavior at low frequency with increasing $\Phi_{si}$, and thus with the volume fraction in the branches, $\Phi_{agg}$. We have therefore adapted a simple percolation model to the reinforcement data in Figure 9. Our model is based on a hydrodynamic description below a critical percolation volume fraction $\Phi_{agg}^c$, and on a percolation expression above [74, 75]

$$\frac{G}{G_0} = 1 + 2.5\ \Phi_{agg}\ + \Theta\Big(\Phi_{agg} - \Phi_{agg}^c\Big)\ \frac{G_f}{G_0}\Bigg(\frac{\Phi_{agg} - \Phi_{agg}^c}{1 - \Phi_{agg}^c}\Bigg)^{b} \qquad (13)$$

where $\Theta(\Phi_{agg} - \Phi_{agg}^c)$ denotes the Heaviside step function (zero for negative arguments, one for positive ones), and $G_f$ is the modulus of the fractal network. Note that eq.(13) relies on eq. (12), which relates the aggregate volume fraction in the branches to the silica volume fraction in a non linear manner. For $\Phi_{agg}$, we have used linear interpolations of the aggregate compacity $\kappa$ and the volume fraction of fractal branches $\Phi_{fract}$ as determined by TEM and SAXS (Table 1). The exponent of the percolation term, b, was set to 1.8 in agreement with the literature [74]. We are thus left with two virtually independent parameters, $\Phi_{agg}^c$ and $G_f/G_0$, which have been varied to optimize the fit. The critical percolation volume fraction of aggregates in the branches is found to be $\Phi_{agg}^{c} = 56\%\,v$, which corresponds to a silica volume fraction of $\Phi_{si}^{c} = 12\%\,v$. The rather high value of $\Phi_{agg}^{c}$ is consistent with our picture of aggregates percolating within the fractal branches, i.e., in a space of reduced dimension. In one dimension, the exact result is a percolation only at full coverage.[76] The remaining parameter is the ratio of the moduli. A value of $G_f/G_0 = 50$ is found to correctly reproduce the increase of the reinforcement factor with silica volume fraction.

Given the simplicity of the rheological model, the compatibility with our previous analysis by SAXS (see in Table 1, $\kappa = 31\% - 38\%$) is encouraging. This underlines the consistency of the methods. In particular, we have checked that fixing the compacity to other values (30% or 40%) reduces the quality of the fit strongly. The ratio of the moduli seems a bit low, as one might expect much higher moduli for pure silica, at least $10^3$ times higher than the one of the matrix. The branches, however, are made of non compact aggregates, with coating agents on the silica nanoparticle surface. These may be the reasons for a lower modulus of the branches. The resulting percolation upturn observed in Figure 9 is thus weaker than in cases of uncoated silica[35], as also observed by Chevigny et al [15], but with a similar filler connectivity threshold. To finish this discussion, one may note that the data could also be described with other models (however with a much lower quality of the fit), like an exponential increase with the filler volume fraction which was found to describe reinforcement data in carbon black reinforced rubbers.[77]

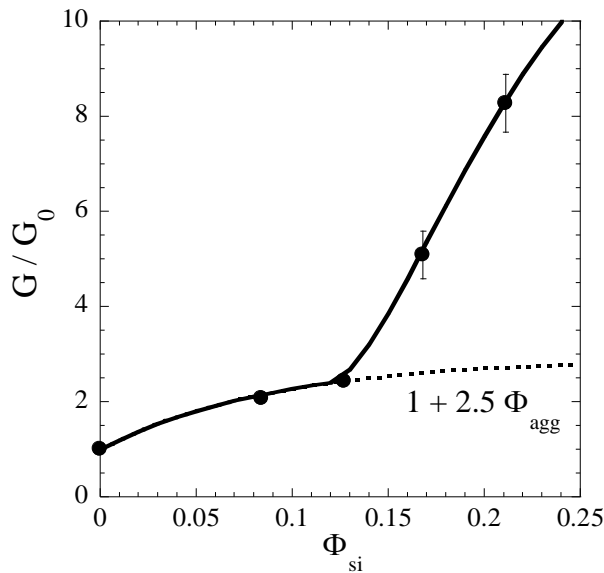

**Figure 9:** Reinforcement factor $G/G_0$ of nanocomposites with $\Phi_{si}$ = 8.4% v - 21.1% v, where G is the storage modulus at 150 Hz, and $G_0$ the corresponding one of the matrix (squares). Line is a fit with eq. (13) using the compacity and $\Phi_{fract}$ of the structural analysis as input. The fit parameters are $\Phi_{fract}^c$ = 0.56 (corresponding to $\Phi_{si}^c$ = 0.12) and $G_f/G_0$ = 50. The purely hydrodynamic reinforcement is also indicated (dotted line).

## 4. Conclusion

The structure of nanocomposites designed to reproduce key features of industrial samples, but of simplified composition, has been studied on length scales extending from the nanometric primary particles to microns. We have developed an original method for scattering data analysis of such multi-scale systems. The combination of TEM, SAXS, and computer simulations allowed for a quantitative analysis, evidencing the formation of small aggregates of average radius in the 35 – 40 nm range, with a large polydispersity in aggregate size (estimated to be about 30%) and thus in aggregation number: most of the aggregates contain some fifteen primary particles, but the average amounts to about forty-five. Compacity of aggregates was assumed to be identical for all sizes, and it was found to increase from 31% to 38% with $\Phi_{si}$. Here one may add that these numbers are necessarily model dependent, which may impact the evolution of the compacity, which in any event stays in the 35%-range. Within our model, we have considered that these aggregates possess excluded volume interactions, which generate a visible shoulder in the scattering curves. It is important to recognize that this shoulder cannot be interpreted as a Guinier-signature of objects. The polydisperse aggregates fill up branches with a volume fraction of aggregates $\Phi_{agg}$ increasing from about 45 to 70%, as $\Phi_{si}$ goes up from 8.4 to 21.1%v. The approximate lateral dimension of the branches is 150 nm, i.e., it is only a few aggregates wide. The large-scale spatial arrangement of the branches can be described by a fractal of average dimension of 2.4. The structure contains pure polymer zones. Their volume fraction $(1-\Phi_{fract})$ decreases from 41 to 20% for 8.4%v and 21.1%v of silica, respectively. To summarize, it is demonstrated that the complex structure of interacting aggregates in nanocomposites of industrial origin can be quantitatively modeled by including self-consistent polydisperse form and structure factors of aggregates.

The rheology of the simplified industrial nanocomposites has been studied as a function of filler volume fraction, in small amplitude oscillatory shear experiments. Master curves for the storage and loss moduli could be constructed up to $\Phi_{si} = 12.7\%v$. These curves display a crossover from a flow regime to solid-like behaviour with increasing filler fraction at low frequency, as well as an increase of the high-frequency moduli. The resulting reinforcement curve of the high-frequency storage modulus can be described using a combination of hydrodynamic reinforcement for $\Phi_{si}$ below a critical percolation volume fraction ($\Phi_{si}^c = 12\%v$), and a percolation law above. It is interesting to note that the aggregate compacity obtained from the structural analysis (SAXS and TEM) is fully compatible with the reinforcement data.

To finish the conclusions of this article, one may note that the polymer matrix was a mixture of reactive and inert chains. The influence of the ratio of reactive chains on the structure will be studied in a forthcoming article [78]. Up to here, following our idea of simplification of the system, we have also deliberately avoided another key ingredient, the coupling agent. Its influence on microstructure in these systems is currently under investigation [79]. Finally, for future work, it may be important to be able to compare the results obtained here to model systems where the filler is a well-defined nanoparticle.

**Acknowledgements**: We are thankful for a "Chercheur d'Avenir" grant by the Languedoc-Roussillon region (JO) and PhD funding "CIFRE" (GB). We are indebted to C. Negrell-Guirao (ENSCM, Montpellier) for experimental support. Fruitful discussions with A. Bouty, F. Boué, J. Jestin (LLB Saclay), and M. In (L2C) are gratefully acknowledged.

## Appendix 1: Typical formulations of industrial nanocomposite systems

| Function | Name | Abbreviation | Simplified system |
|---|---|---|---|
| Coupling agent | bis (3-triethoxysilylpropyl) tetrasulfide | TESPT (Si69) | |
| Coupling agent | 3-mercaptopropyltriethoxysilane reacted with ethoxylated $C_{13}$-alcohol | Si363 | |
| Coating agent | octyl-triethoxysilane | octeo | **X** |
| Catalyzer | diphenyl guanidine | DPG | **X** |
| Cross-linking agent | sulphur | | |
| Cure activator | ZnO particles | | |
| Cure activator | stearic acids | | |
| Cure accelerator | N-butyl-2-benzothiazole sulfonamide | TBBS | |
| Cure accelerator | N-cyclohexyl-2-benzothiazole sulfonamide | CBS | |
| Antioxidant | N-isopropyl-N'-phenyl-para-phenylenediamine | IPPD | |
| Antioxidant | N-(1,3-dimethylbutyl)-N'-phenyl-para-phenylenediamine | 6PPD | **X** |
| Antioxidant | 2,2'-methylenebis-(4-methyl-6-tertiary-butylphenol) | AO2246 | **X** |

**Table A1:** Typical industrial formulations in SBR-silica nanocomposites. The last column indicates the components used in the simplified system studied in this article.

**FOR TABLE OF CONTENTS USE ONLY**

**Multi-scale filler structure in simplified industrial nanocomposite systems silica/SBR studied by SAXS and TEM**

Guilhem P. Baeza, Anne-Caroline Genix, Christophe Degrandcourt, Laurent Petitjean, Jérémie Gummel, Marc Couty, Julian Oberdisse

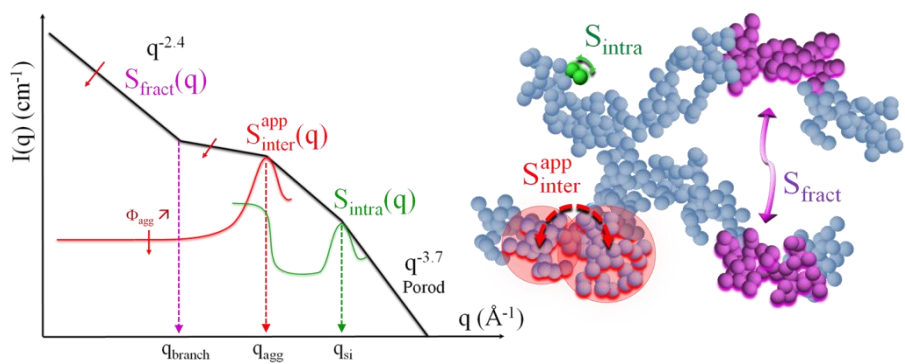